\documentclass[prl,aps,amsfonts,twocolumn]{revtex4}
\usepackage{graphicx}

\begin{document}

\title{Topological Quantum Computing with Only One Mobile Quasiparticle}

\author{S. H. Simon$^1$, N.E. Bonesteel$^2$, M. H. Freedman$^3$,  N. Petrovic$^1$, L. Hormozi$^2$}

\address{
 ${}^1$ Bell Laboratories, Lucent Technologies, 700 Mountain
 Avenue, Murray Hill, New Jersey 07974, USA
 \\ ${}^2$ Department of Physics and NHMFL, Florida State University,
 Tallahassee, Florida 32310, USA
 \\ ${}^3$ Microsoft Research, One Microsoft Way, Redmond, Washington 98052, USA}

\begin{abstract}
In a topological quantum computer, universal quantum computation
is performed by dragging quasiparticle excitations of certain two
dimensional systems around each other to form braids of their
world lines in 2+1 dimensional space-time.   In this paper we show
that any such quantum computation that can be done by braiding $n$
identical quasiparticles can also be done by moving a single
quasiparticle around $n-1$ other identical quasiparticles whose
positions remain fixed.
\end{abstract}

\maketitle

A remarkable recent theoretical advance in quantum computation is
the idea of topological computation
\cite{kitaev,freedman,ogburn,mochon1,mochon2,preskill,us}. Using
exotic two dimensional quantum systems, including certain
fractional quantum Hall states \cite{read,slingerland,pan},
rotating bose condensates \cite{cooper}, and certain spin systems
\cite{nayak,fendley}, it has been shown \cite{kitaev,freedman}
that universal quantum computations can be performed by simply
dragging identical quasiparticle excitations around each other to
form particular braids in the quasiparticles' world-lines in 2+1
dimensions. Because the resulting quantum gate operations depend
only on the topology of the braids formed by these world-lines,
the computation is intrinsically protected from decoherence due to
small perturbations to the system. Realization of such a
topological quantum computer has previously appeared prohibitively
difficult in part because one would have to be able to manipulate
many quasiparticles individually so as to braid them around each
other in arbitrary patterns. In this paper we show that universal
quantum computation is possible using a very restricted subset of
braid topologies (``weaves") where only a single quasiparticle
moves and all the other identical quasiparticles remain
stationary. This simplification may greatly reduce the
technological difficulty in realizing topological quantum
computation.

We note that there are several different proposed schemes for
topological quantum computation
\cite{kitaev,freedman,ogburn,mochon1,mochon2}. In this paper we
will focus on systems of the so-called Chern-Simons-Witten type
\cite{freedman,preskill}. In these systems the topological
properties are described by a gauge group and a ``level" $k$ which
we write as a subscript.  The cases of $SU(2)_k$ are known to
correspond to the properties of certain quantum Hall states
\cite{read,slingerland}.  The $SU(2)_3$ case, which is thought to
have been observed experimentally \cite{read,pan}, is the simplest
such model capable of universal quantum computation
\cite{freedman} and is very closely related to the Fibonacci anyon
model, $SO(3)_3$ \cite{preskill,fendley}. It may also be possible
to realize theories of this type in rotating Bose condensates
\cite{cooper} and quantum spin systems \cite{fendley,nayak}.

The braid group $B_n$ on $n$ strands is a group generated by the
$(n-1)$ elements $\tau_1$ through $\tau_{n-1}$ and their inverses.
As shown in Fig.~1, The generator $\tau_p$ switches the strand at
position number $p$ with the strand at position $p+1$ in a
clockwise manner, whereas the inverse $\tau_p^{-1}$ switches these
strands in a counterclockwise manner (we count strand positions
from bottom to top). By multiplying these generators, any braid on
$n$ strands can be built (see Fig.~1).   Reading an expression
left to right such as $\tau_3 \tau_2 \tau_3^{-1}$ means one should
do $\tau_3$ first followed by $\tau_2$ followed by $\tau_3^{-1}$.
We thus express a general braid as
\begin{equation}
\label{eq:b1}
    \tau^{r(1)}_{s(1)} \, \tau^{r(2)}_{s(2)} \, \tau^{r(3)}_{s(3)}  \, \ldots \, \tau^{r(p)}_{s(p)}
\end{equation}
with $p$ the total number of generators required to express the
braid.  Here each $s(i)$ takes a value in $1 \ldots n-1$ and each
$r(i)$ is either $\pm 1$.

A subset of the braid group $B_n$ is the set of all braids that
move only a single strand (the ``warp" strand in the nomenclature
of weaving) around $n-1$ stationary strands (the ``weft").  We
will call this subset ``weaves" with $n-1$ weft strands. An
example of a weave is shown in Fig.~2.   A braid that is a
non-weave is shown in Fig.~1.

\begin{figure}[t]
\includegraphics[width=.75\columnwidth]{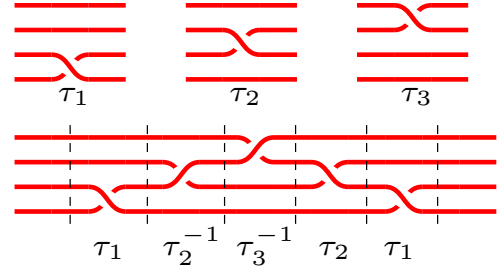}
 \caption{Braids on four strands. {\bf Top:} The three braid
generators.
 {\bf Bottom:} An arbitrary braid on four strands can
be made by multiplying together the generators and their inverses.
This braid shown here, $\tau_1 \, \tau_2^{-1} \tau_3^{-1} \tau_2
\, \tau_1$, is not a weave.}
\end{figure}

\begin{figure}[htbp]\includegraphics[width=.75\columnwidth]{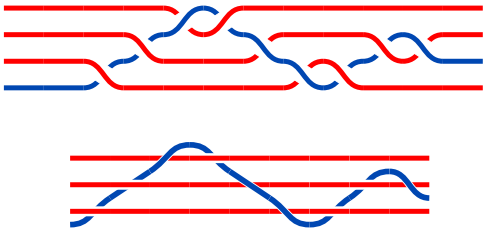}
\caption{The same weave with 3 weft strands drawn two different
ways. In both pictures the warp strand is blue while the weft
strands are red. In the lower picture it is clear that the three
weft strands remain stationary.}
\end{figure}

In topological quantum computation, qubits are encoded in clusters
of quasiparticles. Dragging quasiparticles around each other to
form braids in 2+1 dimensional space-time performs quantum
operations in the Hilbert space of the system. Each braid
generator corresponds to a particular unitary operation, and
performing one braid followed by another corresponds to performing
one quantum operation followed by another. In this way,
complicated gate operations can be built up from simple ones in
the same way that complicated braids are built from the
generators.


To build such a quantum computer it has previously been thought
that one would have to be able to control the motion of $n$
quasiparticles separately (with $n$ proportional to the number of
qubits in the system) such that arbitrary braids can be created.
This amount of control of a (typically microscopic) quantum system
is daunting technologically.  To address this problem, in this
work we will show that the set of weaves is also sufficient to
perform universal quantum computation, and further that such a
weaving computer is ``efficient" in the computational sense. This
result greatly simplifies the challenge of actually building a
topological quantum computer. Now, instead of having to manipulate
$n$ quasiparticles, we need only fix the position of $n-1$ (weft)
quasiparticles and control the motion of a single (warp)
quasiparticle.

By definition, in a topological quantum computer, any quantum
operation on the computational Hilbert space can be approximated
arbitrarily accurately with a braid \cite{freedman}.  We will show
(in part I below) that any quantum operation can also be
approximated arbitrarily accurately with a weave.  Then given any
braid on $n$ quasiparticles made of $p$ generators
(Eq.~\ref{eq:b1}) we show (in part II below) one way to explicitly
construct a weave that performs the same quantum operation as the
given braid to within any desired accuracy $\epsilon$. Further we
show that the particular weave we construct is longer than the
original braid by a factor of at most $C n p |\log
(\epsilon/(np))|^\alpha$ with $C$ a constants depending on the
particular topological theory and $\alpha \simeq 4$. Thus we
demonstrate explicitly that our construction is computationally
``efficient" (since such polynomial and log increases are
acceptable for most quantum computational applications).

{\bf Part I: Dense Image of PureWeaves.}  We define the group
$PB_n$, known as the ``purebraid group on $n$ strands," to be the
subgroup of the braid group on $n$ strands, $B_n$, where each
strand begins and ends in the same position.   A subgroup of the
purebraids on $n$ strands is the ``pureweaves" on $n-1$ weft
strands, $PW_{n-1}$. These are analogously the weaves on $n-1$
weft strands where the warp particle begins and ends at the bottom
position.

Given a group $G$ with a subgroup $H$, we say that $H$ is a
``normal'' subgroup of $G$  if for each $g \in G$ and $h \in H$,
we have $g h g^{-1} \in H$. We now show that $PW_{n-1}$ is a
normal subgroup of $PB_n$. Choosing any purebraid $b$ and any
pureweave $w$, we claim that $b w b^{-1}$ is topologically
equivalent to a pureweave (See Fig.~3). To see that this is true,
we erase the warp strand as shown in Fig.~3, so $b$ maps to a
purebraid $b'$ on $n-1$ strands, $w$ maps to the identity on $n-1$
strands, and $b^{-1}$ maps to $b'^{-1}$. Thus $b w b^{-1}$ maps to
$b' b'^{-1}$ meaning that we obtain the identity once we erase the
warp strand. This implies that the original braid $b w b^{-1}$
must have been a pureweave, proving that the pureweaves $PW_{n-1}$
are a normal subgroup of the purebraids $PB_n$.

\begin{figure}
\includegraphics[width=.9\columnwidth]{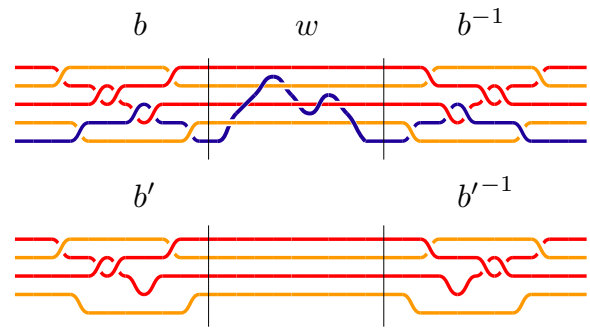}
\caption{Graphical proof that $PW_{n-1}$, the pureweaves with
$n-1$ weft strands, is a normal subgroup of $PB_n$, the purebraids
with $n$ strands.  In the top we construct a purebraid $b$ on 5
strands followed by a pureweave $w$ with 4 weft strands (the warp
strand is blue) followed by the purebraid $b^{-1}$.   To see that
the resulting braid $b w b^{-1}$  is a pureweave, we erase the
warp strand as shown in the bottom. Since the remaining braid is
the identity, $b w b^{-1}$ must have been a weave.}
\end{figure}

We now consider a topological system with $n$ identical
quasiparticles and a Hilbert space of dimension $M$. We assume
that the purebraids $PB_n$ have a dense image in $PU(M)$. Here
$PU(M) = SU(M) /\mathbb{Z}_M$. (The $\mathbb{Z}_M$ subgroup of
$SU(M)$ is generated by $e^{2 \pi i/M}$ times the identity. Since
this is just an overall phase factor, it is irrelevant for quantum
computation). The statement that $PB_n$ has a dense image in
$PU(M)$ means essentially that given an element $a \in PU(M)$
there exists a braid in $PB_n$ corresponding to some element
$\tilde a \in PU(M)$ in the Hilbert space whose value is
arbitrarily close to $a$. This statement is necessarily true if
one can do universal quantum computation with quasiparticles of
the theory (which is what we are assuming).  We note that for
$SU(2)_k$ Chern-Simons-Witten theories it has been shown
\cite{freedman,endnote3} that the purebraids on $n$ strands do
indeed have a dense image in $PU(M)$ for $k>2$ and $k \ne 4,8$
when $n = 3$, and for $k>2$ and $k\ne 4$ when $n > 3$.

Since the group $PB_n$ has a dense image in $PU(M)$, the normal
subgroup $PW_{n-1}$ of $PB_n$ must then have an image which is
dense in some normal subgroup of $PU(M)$.  However, it is a well
known result \cite{freedman} that $PU(M)$ has no normal subgroups
except for the identity and the entire group $PU(M)$ itself. Since
it is easy to show \cite{endnote5} that the pureweaves $PW_{n-1}$
do not all map to the identity, $PW_{n-1}$ must also have a dense
image in all of $PU(M)$.  Thus we have shown that any quantum
operation can be approximated arbitrarily accurately with a weave.

To be more precise about the key piece of this argument we can
state the following {\it Lemma:} If a group $H$ (such as
$PW_{n-1}$) is a normal subgroup of $G$ (such as $PB_n)$ and $G$
is mapped by a group homomorphism $\rho$ densely into a compact
topological group $T$ (such as $PU(M)$) then $S =
\mbox{closure}(\mbox{image}(H) )$ is a normal subgroup of $T$.
{\it Proof:} Assume $S$ is not normal in $T$, so that there must
exist a $t \in T$  such that $ t S t^{-1}$ is some subgroup $S'$
different from $S$.  Since $G$ is mapped densely into $T$, there
must exist a sequence of $t_i \in T$ which converges to $t$ where
each $t_i = \rho(g_i)$ for some $g_i \in G$. The limit of the
sequence $\mbox{closure}(\mbox{image}(g_i H g_i^{-1}) )$ must then
be $\lim [ t_i (\mbox{closure}(\mbox{image}(H) ) t_i^{-1}] = \lim
t_i S t_i^{-1} = t St^{-1} = S'$.  But since $H$ is normal in $G$,
we must have $g_i H g_i^{-1} = H$ for any $g_i$ so each element of
the sequence must give $S$, contradicting our assumption that $S'
\neq S$. (QED).

\begin{figure}[t]\includegraphics[width=.9\columnwidth]{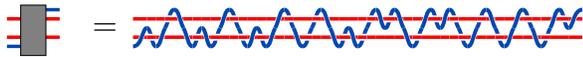}
\caption{An example of an injection weave for the Fibonacci anyon
model ($SO(3)_3$ or $SU(2)_3$).  This injection was first
discussed in Ref.~\onlinecite{us} and approximates the identity
operation on the Hilbert space to better than one part in $10^2$,
while transferring the warp quasiparticle from the bottom to the
top. Longer weaves will approximate the identity exponentially
more closely with the weave becoming longer only linearly
\cite{us,KitaevSolovay}. The box on the left establishes the
notation used in Fig.~5 below.}
\end{figure}

{\bf Part II: Explicit Construction.} Our construction is based on
the ``injection weave" first discussed in Ref.~\onlinecite{us}.
This is a weave on three strands (two weft strands), approximating
the identity operation on the Hilbert space, which starts with the
warp strand as the bottom of the three strands and ends with the
warp strand as the top of the three strands.  We can similarly
define the inverse of the injection weave which moves the warp
from the top to the bottom of the three strands.

The Kitaev-Solovay theorem \cite{KitaevSolovay} along with our
above Lemma \cite{endnote2} guarantee that for any system of
Chern-Simons-Witten type capable of topological quantum
computation it is possible to efficiently find an injection weave
of length $C |\log\epsilon|^\alpha$ where $\epsilon$ is a measure
of the distance of the resulting gate from the identity, $\alpha
\simeq 4$, and $C$ is a constant depending on the particular
topological theory we are considering.  Thus, with linearly
increasing complexity of the injection weave, the identity can be
approximated exponentially more accurately \cite{us,endnote6}.

In Ref.~\onlinecite{us}, examples of injection weaves were
explicitly constructed for Fibonacci anyons
\cite{freedman,preskill}. One such example is shown in Fig.~4.
(The same injection weave applies for the elementary
quasiparticles of the experimentally observed \cite{pan} $SU(2)_3$
system). It is useful to think back to the Fibonacci anyon case as
a concrete example, although our construction is much more
general.

We now consider multiple injections.   Suppose the warp is strand
number $m$ at a given point in time and we would like to move the
warp until it is strand number $m+2q$ (with $q$ an integer)
without disturbing the state of the system. We do this by
repeating the injection weave
\begin{eqnarray}
    & & M_{m;m+2q} =  \\ \nonumber & & \left\{\begin{array}{ll}  I_{m,m+2} I_{m+2,m+4} \ldots
    I_{m+2q-2,m+2q}  &  ~~~~~ q > 0
     \\ I^{-1}_{m-2,m} I^{-1}_{m-4,m-2} \ldots
    I^{-1}_{m+2q,m+2q+2}  &  ~~~~~ q < 0
    \end{array} \right.
\end{eqnarray}
Here, $I_{a,a+2}$ is an injection weave acting on strands $a, a+1$
and $a+2$ where the warp starts at position $a$ and ends at
position $a+2$.   Similarly, $I^{-1}_{a-2,a}$ is an injection
acting on strands $a-2,a-1$ and $a$ which moves the warp from
position $a$ to position $a-2$.   Thus, the multiple injection
$M_{m;m+2q}$ moves the warp from position $m$ to position $m+2q$
while performing only (approximately) the identity operation on
the Hilbert space. Note that $M_{m,m}$ is defined to be the
identity, since no braiding is needed to move the warp from
position $m$ to position $m$.

\begin{figure}[t]
\includegraphics[width=.7\columnwidth]{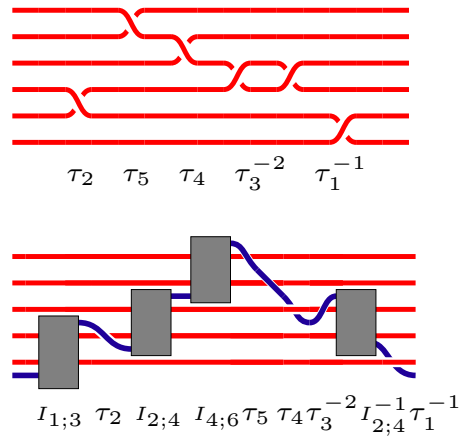}
 \caption{We construct a weave (bottom) which
produces the same quantum operation on the Hilbert space as some
desired arbitrary braid (top).  In the bottom, the shaded boxes
represent injection weaves which have (approximately) no effect on
the Hilbert space. This construction shows that, so long as an
injection weave exists, weaves are just as capable as braids at
performing efficient universal quantum computation.}
\end{figure}

We consider an arbitrary braid expressed as in (\ref{eq:b1})
above. Starting with the warp on the bottom at position 1, we do
multiple injections until the warp is in a position to make the
first desired braid operation $\tau_{s(1)}^{r(1)}$. Defining
$[a]_2 = a \, \rm{mod} \, 2$, our first step is then $M_{1;s(1) -
[s(1)]_2 + 1}$ which performs (approximately) the identity on the
Hilbert space, but moves the warp an even number of strands over,
placing it in position to do the desired $\tau^{r(1)}_{s(1)}$.
After performing $\tau^{r(1)}_{s(1)}$, the warp is occupying an
even numbered position. We then make multiple injections to move
the warp to a position where it can do the next braid operation
$\tau_{s(2)}^{r(2)}$, after which the warp occupies an odd number
position again, and so forth. Generally, let us define
\begin{equation}
    \tilde M(i) = \left\{ \begin{array}{ll} M_{s(i-1)+[s(i-1)]_2 \, ; \, s(i) + [s(i)]_2
    } & \mbox{even } i \\
    M_{s(i-1)-[s(i-1)]_2+1 \,  ; \, s(i) - [s(i)]_2+1 } & \mbox{odd }  i
\end{array} \right.
\end{equation}
Thus, defining $s(0) = 0$, we can write out the full weave that
performs the same quantum operation as the braid written in
expression (\ref{eq:b1}) above
\begin{eqnarray}
    \tilde M(1) \,  \tau_{s(1)}^{r(1)} \,  \tilde
    M(2) \,
    \tau_{s(2)}^{r(2)} \, \tilde M(3) \ldots  \tilde M(p) \,  \tau_{s(p)}^{r(p)}
\end{eqnarray}
Fig.~5 shows this construction graphically.   By construction, to
the extent that $\tilde M$ correctly performs the identity on the
Hilbert space, this weave performs the same operation on the
Hilbert space as any given braid in expression (\ref{eq:b1}). The
constructed weave is longer than the given braid by no more than
$np$ times the length of the needed injection weave.  Further if
we want to approximate the quantum operation of the original braid
to within some accuracy $\epsilon$, each $\tilde M$ need only be
equal to the identity to within $\epsilon/p$. Thus, each injection
weave need only be equivalent to the identity to within
$\epsilon/(np)$ which requires the injection to be length $C |\log
(\epsilon/(np))|^\alpha$.  Note that since the constant $C$ is
determined entirely from the injection weave on three strands, it
is independent of $n$ and $p$. Thus the total length of the
constructed weave need be no longer than $C np|\log
(\epsilon/(np))|^\alpha$ as claimed.   This length estimate should
be viewed as an upper bound and proof of principle.   In fact, we
expect that weaves may be efficiently found which are
significantly shorter than those constructed here. Nonetheless,
the concept of injection can also be used to design more practical
weaves, as will be discussed in forthcoming work.

{\bf Further Comments:} In this work we have nowhere discussed the
initialization or readout steps required for quantum computation.
This is a difficult problem that has not been satisfactorily
answered anywhere in the literature for this type of topological
quantum computer.  It has been proposed that measurements and
initialization could be achieved in principle using interference
experiments \cite{slingerland2,dassarma}, or by fusion of
quasiparticles \cite{kitaev,freedman}. However, the precise
initialization and measurement schemes will depend heavily on the
particular nature of the realization of the computer when it is
built.

The authors acknowledge G. Zikos for helpful conversations, and I.
Berdnikov for his careful proofreading of this manuscript.  We
also thank the referees for forcing us to be more precise about
the issue of efficiency.  NEB and LH acknowledge support from US
DOE Grant DE-FG02-97ER45639.







\begin{references}
\bibitem{endnote3}
Regarding our assumption that the image of $PB_n$ is dense in
$PU(M)$,  if the image of the braid group $B_n$ is dense in
$PU(M)$ then since $PB_n$ is a normal subgroup of $B_n$, and the
image of $PB_n$ is clearly not the identity, then by using the
same Lemma proved in this work, $PB_n$ is also dense in $PU(M)$
for these theories.
\bibitem{endnote5} To show that the image of $PW_{n-1}$ is not the identity,
we assume the converse. This would imply that the square of a
braid generator is the identity. However, when the square of the
braid generator is the identity, the representation of the braid
group must be the finite symmetric group, which therefore
contradicts the assumption that we started with a dense
representation of $PB_n$.
\bibitem{endnote6} For all $SU(2)_k$ models with $k > 2$ we have
numerically found approximate injection weaves similar to those
found for $k=3$ in Ref. \onlinecite{us} and we have shown that
they can be systematically improved to any desired accuracy
\cite{KitaevSolovay}. Note that for $k=4 m$, the weave
$\tau_i^{2m+1}$ provides an exact injection of finite length which
transfers the warp over a single weft particle \cite{slingerland}.
\end{references}
\end{document}